\documentclass{PoS}

\def\bea{\begin{eqnarray}}
\def\eea{\end{eqnarray}}

\title{The model dependence of $m_\rho / f_\pi$}

\ShortTitle{The model dependence of $m_\rho / f_\pi$}

\author{\speaker{Daniel Nogradi} and Lorinc Szikszai\\
        Eotvos University, Department of Theoretical Physics, Budapest 1117, Hungary\\
        E-mail: \email{nogradi@bodri.elte.hu}, \email{szikszail@caesar.elte.hu}}

\abstract{Should a strongly coupled composite Higgs boson scenario be realized in Nature the most
easily accessible experimental signal would be new particles made up of the same ingredients as the Higgs
but with different quantum numbers. The lightest of these hypothetical new particles would probably be the vector mesons.
In this contribution we report results on $m_\rho / f_\pi$ in the chiral-continuum 
limit with $SU(3)$ gauge group and $N_f = 2,3,4,5,6$ flavors of fundamental fermions. In addition we compare $m_\rho /
f_\pi$ results from various models with different gauge groups and fermion content.
The main conclusion seems to
be that the experimental measurement of this vector meson mass will be able to distinguish between gauge groups but less so
between the fermion content.
}

\FullConference{37th International Symposium on Lattice Field Theory - Lattice2019\\
		16-22 June 2019\\
		Wuhan, China}

\begin{document}

\section{Introduction}
\label{intro}

It may very well be the case that the experimentally discovered Higgs boson is the elementary spin-0 particle 
of the Standard Model and the theoretical description is valid up to very high energies beyond the scope of the LHC or
any new accelerator in the near future. If on the other hand the Higgs boson turns out to be a composite particle the
bare minimum experimental signal should be the discovery of new particles made up of the same ingredients as the composite
Higgs itself but with different quantum numbers. The lightest candidates in a large class of strongly interacting
electro-weak models would be the vector mesons in the non-abelian gauge theory responsible for the composite Higgs. The
physical scale in these technicolor inspired class of models is given by the decay constant of the Goldstone bosons
$f_\pi = 246\;GeV$. Lattice calculations are ideal to obtain the massless ratio $m_\rho / f_\pi$ in the chiral-continuum
limit from first principles. Conceptually it is a simple task: pick a gauge group $G$, representation $R$ and flavor number
$N_f$ (or potentially several representations and several flavor numbers if not all fermions transform in the same
representation), make sure that finite volume effects are small, make sure $\rho$-decay is handled properly, obtain
$a m_\rho$ and $a f_\pi$ in lattice units for various lattice spacings and fermion masses and finally perform a
chiral-continuum extrapolation to obtain the ratio. Thus traditional meson spectroscopy will lead to $m_\rho / f_\pi$ for
every choice $(G,R,N_f)$. If experiments do detect a new particle in the $TeV$ range the next task would be to find
which $(G,R,N_f)$ combination(s) may give rise to that particular mass and quantum number hence it is important to have
scanned as many models as possible \footnote{Throughout this contribution the normalization of the decay constant is
such that $f_\pi = 93\;MeV$ in QCD.}.

Recent lattice studies of the meson spectrum with $SU(3)$ indicate that the fermion content dependence might be very
mild for the ratio $m_\rho / f_\pi$. 
Even though many of the results are at finite lattice spacing, occasionally finite fermion mass and finite volume
effects are not always fully controlled the indications are $m_\rho / f_\pi \sim 8.0$ from a broad range of studies
\cite{Fodor:2009wk, Jin:2009mc,Fodor:2012ty,  Aoki:2013xza, Jin:2013hpa, Fleming:2013tra, Fodor:2016pls, Fodor:2016wal,
Appelquist:2016viq, Appelquist:2018yqe}. As long as the gauge group is fixed, it seems then, that the experimental
measurement of $m_\rho$ will not be able to distinguish the various models as long as the gauge group is $SU(3)$.

In this contribution we report fully controlled simulation results in the infinite volume, chiral and continuum limits
with $SU(3)$ and $N_f = 2,3,4,5,6$ reinforcing the above
indications \cite{1905.01909}. We obtain $m_\rho / f_\pi = 7.95(15)$ and no statistically significant $N_f$-dependence. 

The ratio $m_\rho / f_\pi$ however is expected to change rapidly with the gauge group based on large-$N$ arguments. More
precisely if the fermionic degrees of freedom scale as $O(N)$ which is the case with fundamental fermions, we have
$m_\rho \sim O(1)$ and $f_\pi \sim O(\sqrt{N})$ hence the natural combination to consider is $\sqrt{N}\; m_\rho / f_\pi$.
In the large-$N$ limit this combination is finite. Motivated by this we will be comparing various models via the
combination $\sqrt{dim(R)} \; m_\rho / f_\pi$ where $dim(R)$ is the dimension of the representation; see
\cite{1304.4437, Castagnini:2015ejr, DeGrand:2016pur} and references therein.

In the next section the details of the $SU(3)$ calculations with $N_f = 2,3,4,5,6$ fundamental fermions will be given.
In section \ref{dependence} we will compare results from the literature for various $(G,R,N_f)$ by collecting $\sqrt{dim(R)}\;
m_\rho / f_\pi$ if possible in the chiral-continuum limit, or at finite lattice spacing and/or finite fermion mass if 
chiral-continuum limit results are not available. In section \ref{conclusion} we end with notes on possible future studies.

\section{Results with $SU(3)$ and fundamental fermions}
\label{results}

The lattice discretization is via the tree level improved Symanzik gauge action and stout improved staggered fermions
with 4 steps of stout smearing and $\rho = 0.12$ smearing parameter \cite{Morningstar:2003gk}. 
Ensembles are generated at four lattice spacings
with four fermion masses at each, for every $N_f$. The parameters were chosen such that the $\rho$ is stable and the
finite volume corrections from Luscher's formalism  \cite{Luscher:1990ux, Luscher:1991cf}
predict small finite volume effects. Furthermore, exponential finite volume
effects are ensured to be small as well by having $m_\pi L > 3.10 + 0.35 N_f$ which was shown to lead to less than $1\%$ 
corrections in low energy observables for $2 \leq N_f \leq 6$. 

\begin{figure}
\begin{center}
\includegraphics[width=7cm]{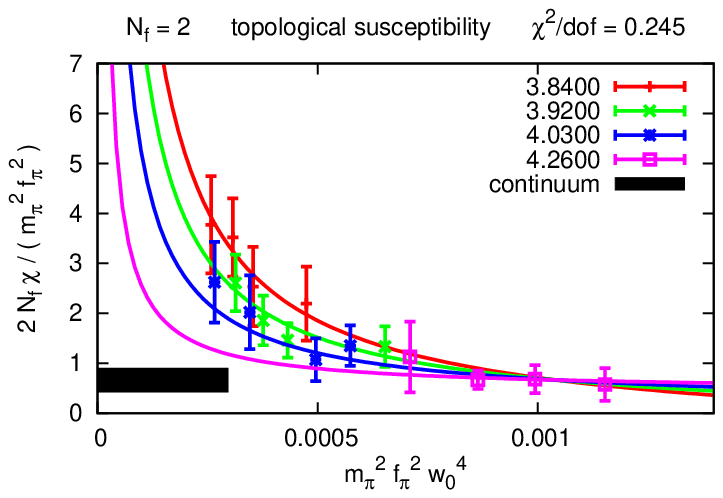} 
\includegraphics[width=7cm]{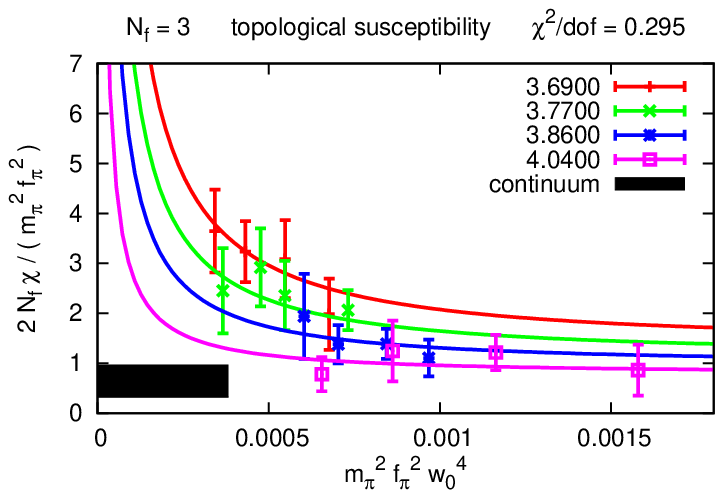} 
\includegraphics[width=7cm]{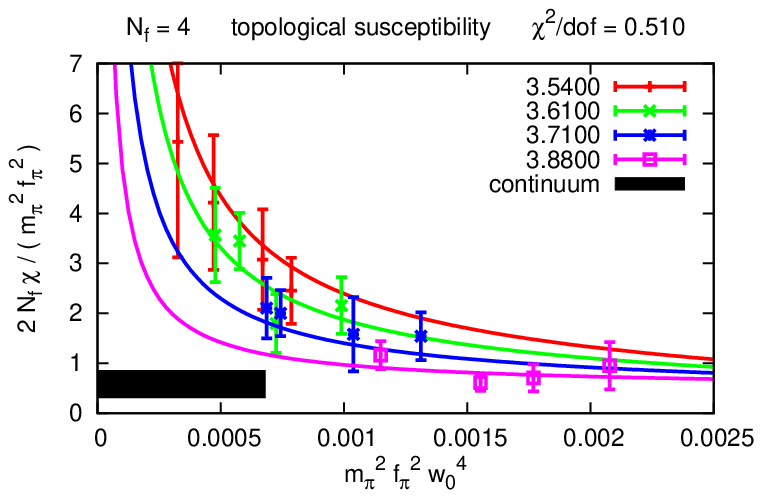} 
\includegraphics[width=7cm]{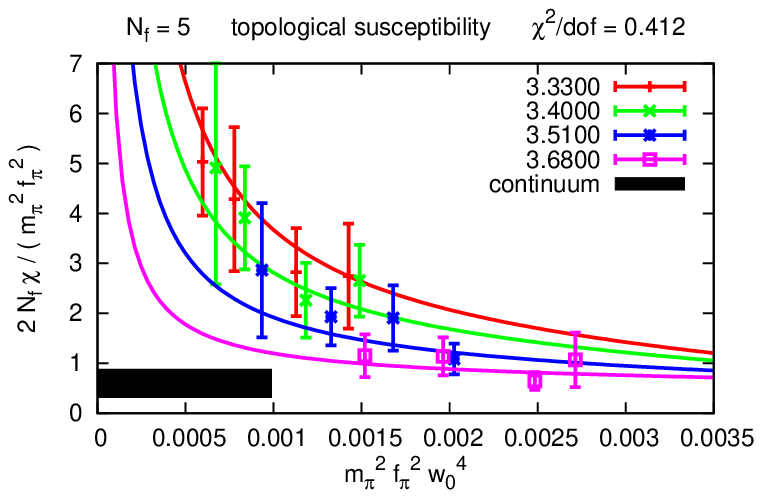} 
\includegraphics[width=7cm]{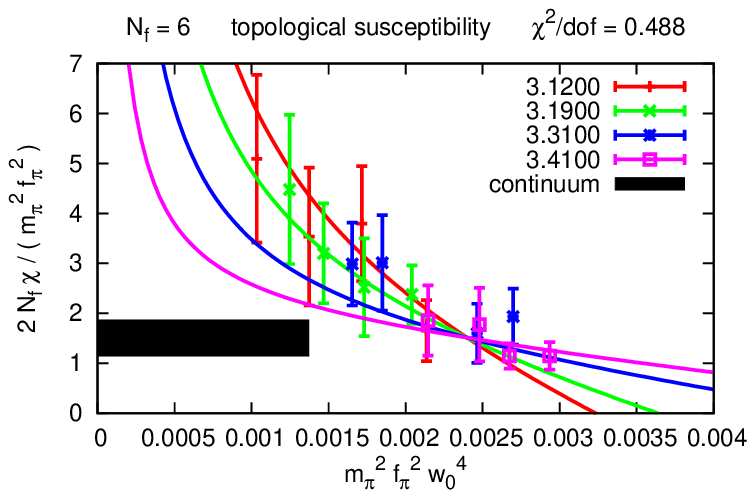} 
\end{center}
\caption{
Chiral-continuum extrapolation of the topological susceptibility. The ratio $2 N_f \chi / ( m_\pi^2 f_\pi^2)$
is shown which, at leading order of chiral perturbation theory, is expected to be constant $1$. The shaded black region
is the result of the chiral-continuum extrapolation. The various colors correspond to various lattice spacings labeled
by the bare coupling $\beta$ in the legend.}
\label{finvolplottopsusc}
\end{figure}

Topology change is frequent enough in all runs so that the topological susceptibility can be
measured accurately enough. In fact the topological susceptibility was used to test for taste breaking effects via the
tree level relation \cite{Leutwyler:1992yt},
\bea
\label{chi}
\frac{\langle Q^2 \rangle}{V} = \chi = \frac{1}{2N_f} f_\pi^2 m_\pi^2 + \ldots\;.
\eea
More precisely, $f_\pi^2 m_\pi^2 / ( 2 \chi )$ can be considered an effective flavor number and if in the
chiral-continuum limit the actual flavor number is obtained within errors, one can be sure that the taste broken
Goldstones of the staggered formulation are scaled to zero as $O(a^2)$ as expected
\cite{Billeter:2004wx}. This test is far easier to perform
than the full spectroscopy of all taste broken Goldstones directly although relatively long runs are required as the
autocorrelation time of the topological charge is typically longer than other quantities. 
At the same time the topological susceptibility
needs to be monitored anyway in order to avoid frozen topology simulations. More precisely, the chiral-continuum
extrapolation is via,
\bea
\label{chifit}
\chi w_0^4 = C_0  m_\pi^2 f_\pi^2 w_0^4  + C_1 \frac{a^2}{w_0^2} + C_2 \frac{a^2}{w_0^2} ( m_\pi^2 f_\pi^2 w_0^4 )\;,
\eea
where $w_0$ is the gradient flow scale \cite{Borsanyi:2012zs}.
In figure \ref{finvolplottopsusc} the
quantity $2N_f\chi / ( m_\pi^2 f_\pi^2 )$ is shown for each $N_f$ for all lattice spacings and masses, together with the
chiral-continuum limit, which ought to be consistent with $1$.
The deviation from $1$ is at most $1.5\,\sigma$.

The 16 simulation points at each $N_f$ allow for controlled chiral-continuum fits of the quantities $f_\pi$ and $m_\rho$
as well. The 4 parameter fits we use are,
\bea
X w_0 = C_0 + C_1 m_\pi^2 w_0^2 + C_2 \frac{a^2}{w_0^2} + C_3 \frac{a^2}{w_0^2} m_\pi^2 w_0^2
\eea
where $X$ is either $m_\rho$ or $f_\pi$. 
Each fit has thus 12 degrees of freedom. The statistical uncertainty of $f_\pi$
is negligible compared to $m_\rho$ and the latter will dominate the final result for the ratio $m_\rho / f_\pi$. With
our action both cut-off effects and finite fermion mass effects are rather small for the ratio \cite{1905.01909}.

The final results are shown in figure \ref{comp} as the left most set of points. Interestingly, a constant fit leads to
$m_\rho / f_\pi = 7.95(15)$ with $\chi^2/dof = 0.26$ i.e. the observed $N_f$-dependence is statistically insignificant. 

\section{Dependence on the gauge group, representation and flavor number}
\label{dependence}

\begin{figure}
\hspace{-2.5cm} \includegraphics[width=20cm]{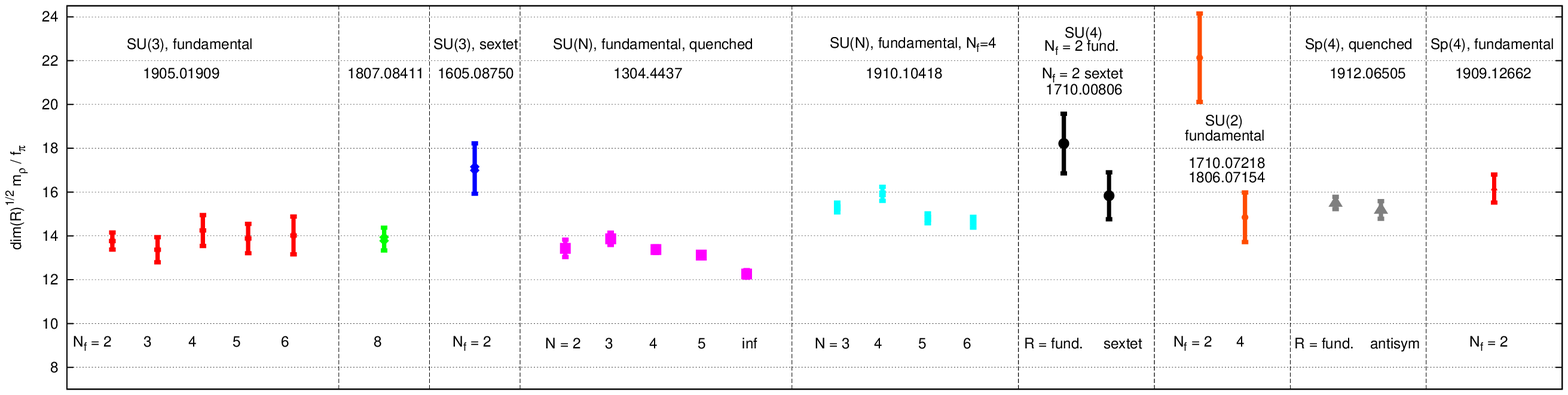}

\hspace{-2.5cm} \includegraphics[width=20cm]{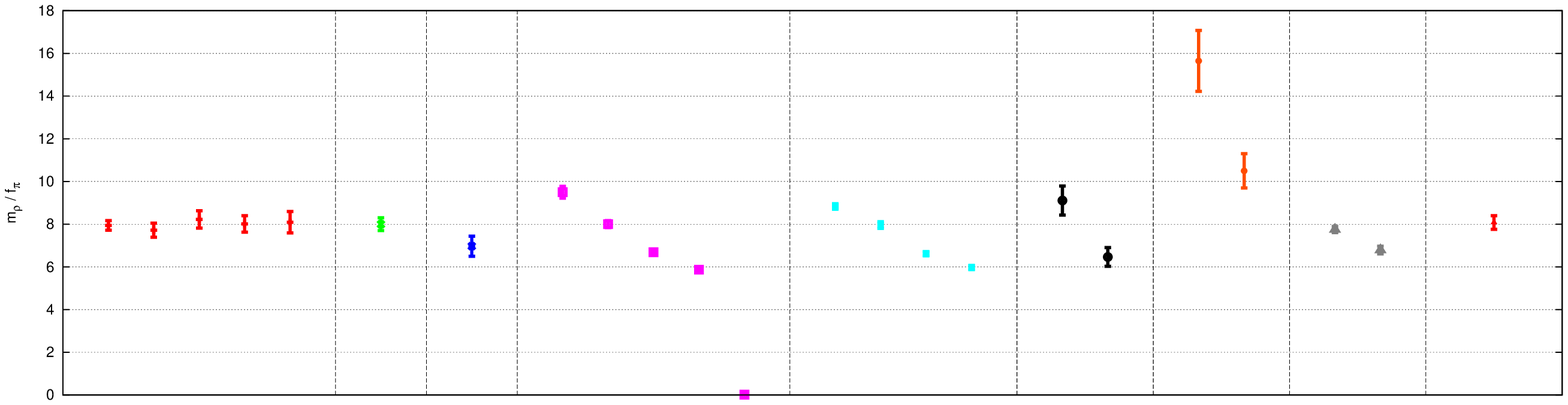}
\caption{Top: comparison of $\sqrt{dim(R)}\;m_\rho / f_\pi$ for various models. Bottom: similar comparison of $m_\rho /
f_\pi$ directly, the labeling of the models is the same as at the top. The $SU(3)$ fundamental $N_f = 2,3,4,5,6$ are in
the chiral-continuum limit \cite{1905.01909}, the $N_f = 8$ result is at finite lattice spacing and finite fermion mass
\cite{Appelquist:2018yqe}, the $SU(3)$ sextet $N_f = 2$ is at finite lattice spacing and finite fermion mass
\cite{Fodor:2016pls},
the $SU(N)$ fundamental quenched results are at finite lattice spacing 
in the chiral limit \cite{1304.4437}, the
$SU(N)$ fundamental $N_f = 4$ are at finite lattice spacing and finite fermion mass \cite{1910.10418, 1907.11511}, 
the $SU(4)$ fundamental $N_f = 2$,
sextet $N_f = 2$ (both are in the sea simultaneously) results are in the chiral-continuum limit \cite{1710.00806}, 
the $SU(2)$ fundamental $N_f = 2$ is in
the chiral-continuum limit \cite{1602.06559, Drach:2017btk}, 
the $SU(2)$ fundamental $N_f = 4$ is at finite lattice spacing and finite fermion mass \cite{1806.07154} and
finally with $Sp(4)$ both the quenched \cite{Bennett:2019cxd} and the fundamental $N_f = 2$ are in the chiral-continuum limit 
\cite{1712.04220, 1909.12662}. Results at finite lattice spacing and/or finite fermion mass should be interpreted with
caution as they naturally contain further systematic errors.}
\label{comp}
\end{figure}

Quite a large number of models were studied recently, partially or completely motivated by a strongly coupled composite
Higgs scenario. Another set of motivations useful for our purposes is the study of the systematics of the
large-$N$ expansion because in this context the gauge group dependence can be displayed. The large-$N$ limit is instructive
because the observed pattern at $SU(3)$ namely no statistically significant dependence on $N_f$ is in line with large-$N$
expectations. As long as the fermionic degrees of freedom scale with $O(N)$, which is the case in the fundamental, one
expects exactly no dependence on $N_f$ in the large-$N$ limit. With the fundamental representation $f_\pi$ scales as
$\sqrt{N}$. If the representation is other than the fundamental it is natural to consider $\sqrt{dim(R)}\; m_\rho / f_\pi$
where $dim(R)$ is the dimension of the representation. If the fermionic degrees of freedom scale as $O(N^2)$ then
of course the usual large-$N$ arguments do not apply.

Starting from $SU(2)$, chiral-continuum results are available with $N_f = 2$ fermions in the fundamental representation
\cite{1602.06559, Drach:2017btk},
$N_f = 4$ at finite lattice spacing and fermion mass \cite{1806.07154}, 
and of course the pure gauge case. The aforementioned large-$N$ studies 
led to results with $SU(N)$, still in the pure gauge 
case, with $N = 2, 3, 4, 5, 6, 7, 17$ and the $N = \infty$ in the chiral limit, at finite lattice spacing 
\cite{1304.4437, Castagnini:2015ejr}. 
Note that with $SU(2)$ all irreducible representations
are real. Increasing the gauge group to $SU(3)$ we have of course QCD results (or Nature) and the $N_f =
2,3,4,5,6$ results of our work in the chiral-continuum limit \cite{1905.01909} as well as $N_f = 8$ at finite
lattice spacing and finite fermion mass \cite{Appelquist:2018yqe}. Still with $SU(3)$ results are available with $N_f =
2$ sextet fermions at finite lattice spacing and finite fermion mass.
Further, $SU(4)$ was studied with two species of fermions simultaneously in
the sea, $N_f = 2$ fundamental together with $N_f = 2$ sextet (which is real) \cite{1710.00806}; see also \cite{1501.05665}
for results with $N_f = 2$ sextet fermions only. 
In this case the two species of fermions
lead to two different decay constants and two different vector meson masses. Results are available in the
chiral-continuum limit \cite{1710.00806}. 
The large-$N$ limit was investigated with $N_f = 4$ fundamental fermions and $N = 3, 4, 5, 6$, the
results are currently at finite lattice spacing and finite fermion mass \cite{1910.10418, 1907.11511}. 
Going beyond the unitary groups, the symplectic
group $Sp(4)$ (which is sometimes denoted by $Sp(2)$, in any case it is the double covering group of $SO(5)$) 
was investigated
in the quenched case with both fundamental and 2-index-antisymmetric representation fermions \cite{Bennett:2019cxd}
and $N_f = 2$ fundamental fermions \cite{1712.04220, 1909.12662}; all the $Sp(4)$ results are in the chiral-continuum
limit.

A summary of all of these results are shown in figure \ref{comp} via the scaled ratio $\sqrt{dim(R)}\; m_\rho / f_\pi$
in the top panel.
In order to better display the dependence on the gauge group the unscaled $m_\rho / f_\pi$ is also shown in the
bottom panel of figure \ref{comp}.
Clearly, the dependence on the fermion content is much less pronounced than the dependence on the gauge
group. However when the leading gauge group dependence is factored out, i.e. $\sqrt{dim(R)}\; m_\rho / f_\pi$ is
considered, even the gauge group dependence is rather small. Only $SU(2)$ shows considerable deviation from large-$N$
scaling and large $N_f$-dependence (although the $N_f = 4$ results are at finite lattice spacing and finite fermion
mass). Interpretation of results at finite fermion mass and/or finite lattice spacing should be done with care of
course.

If one envisions an experimental signal for $m_\rho$ the gauge group can perhaps be narrowed down using figure
\ref{comp} however the information on the fermion content must come from other experimentally accessible quantities
unless the gauge group is $SU(2)$. 

\section{Conclusion and outlook}
\label{conclusion}

A strongly interacting composite Higgs boson is clearly an exciting possibility. The best case scenario would be the
detection of new particles in the TeV range by the LHC. In order to make full use of this result it would be desirable
to obtain fully controlled (volume, lattice spacing, fermion mass) lattice results on the masses of the new particles 
for as many promising models as
possible. Currently the set of fully controlled results is rather limited and even where they are available often
the errors are too large for an experimental result to differentiate between models. It does not seem out of reach to
obtain fully controlled results with $\sim 3-5\%$ errors with $SU(2)$ and $N_f =2,3,4$ and possibly $N_f = 5$
fundamental fermions if the latter is chirally broken. With $SU(3)$ and fundamental fermions $N_f = 7,8,9,10$ seems
possible and it is conceivable that the $Sp(4)$ with $N_f = 2$ results \cite{1712.04220, 1909.12662} can 
be extended to $N_f = 3,4$. The closer a model is to the conformal window the more difficult the calculation becomes
because of large systematic uncertainties. This is the reason why $SU(2)$ with $N_f = 5, 6$ is difficult and so is
$SU(3)$ with $N_f = 2$ sextet \cite{Fodor:2012ty, Fodor:2016pls}. 
Fixing $N_f$ and increasing $N$ on the other hand is moving the model 
away from the conformal window so should be relatively straightforward. In this direction $m_\rho / f_\pi$ is decreasing
and can be arbitrarily small for large enough $N$. Presumably the largest value will occur for $SU(2)$ although 
this is currently an open question.

\section*{Acknowledgements}

We would like to thank William Jay, Julius Kuti, Jong-Wan Lee, Fernando Romero Lopez, Biagio Lucini, Ben Svetitsky and 
Chik Him Wong for sharing data on $m_\rho / f_\pi$ in the various models presented in figure \ref{comp}.
This work was supported by the NKFIH under the grant 
KKP-126769. We are grateful to Kalman Szabo for code sharing.

\end{document}